\documentclass{ws-procs9x6}

\begin{document}

\title{MAPPING STRING STATES INTO PARTONS:\\ FORM FACTORS AND THE
  HADRON SPECTRUM IN AdS/QCD}

\author{G. F. DE T\'ERAMOND$^*$}

\address{Universidad de Costa Rica,\\
San Jos\'e, Costa Rica\\
$^*$E-mail: gdt@asterix.crnet.cr\\}

\begin{abstract}
New developments in holographic QCD are described in this talk in the
context of the correspondence between string states in AdS and 
light-front wavefunctions of hadronic states in physical space-time.
\end{abstract}


\bodymatter


\vspace{10pt}

The AdS/CFT correspondence\cite{Maldacena:1997re} gives unexpected
connections between seemingly different theories which
represent the same observables. On the bulk side it describes the
propagation of weakly coupled
strings, where physical quantities are computed using an effective
gravity approximation. The duality provides a non-perturbative
definition of quantum gravity in a ($d+1$)-dimensional AdS
spacetime in terms of a
$d$-dimensional conformally-invariant quantum field theory at the anti--de Sitter (AdS) 
boundary\cite{Gubser:1998bc}.

The AdS/CFT duality has the potential for understanding fundamental properties of quantum 
chromodynamics such as confinement and chiral symmetry breaking which
are inherently non-perturbative.  As shown by  Polchinski
and Strassler\cite{Polchinski:2001tt}, the AdS/CFT duality, modified
to incorporate a mass scale, 
provides a non-perturbative derivation of dimensional counting
rules\cite{Brodsky:1973kr} for the leading 
power-law fall-off of hard scattering. 
The modified theory generates the hard behavior expected from QCD, instead of the soft
behavior characteristic of strings. 

In its original formulation\cite{Maldacena:1997re}, a correspondence was
established between the supergravity approximation to Type IIB string theory
on a curved background asymptotic to the product space of AdS$_5 \times S^5$ 
and the large $N_C$, $\mathcal{N} = 4$,
super Yang-Mills (SYM) gauge theory in four dimensional space-time. 
The group of conformal transformations
$SO(2,4)$ which acts at the AdS boundary, is 
isomorphic with the group of isometries of AdS space, 
and $S^5$
corresponds to the $SU(4) \sim SO(6)$ global symmetry which rotates the particles
present in the SYM supermultiplet.  The supergravity duality requires
a large AdS radius $R$ corresponding to a large value of the 't Hooft
parameter $g_s N_C$, where $R = ({4 \pi g_s N_C})^{1/4} \alpha_s'^{1/2}$ and
$\alpha_s'^{1/2}$ is the string scale. The classical  approximation
corresponds to the stiff limit where the
string tension $T = R^2 / 2 \pi \alpha' \to \infty$, effectively
suppressing string fluctuations.

QCD is fundamentally different from SYM theories where all the
matter fields transform in adjoint multiplets of $SU(N_C)$.
QCD is also a confining theory in the infrared
with a mass gap $\Lambda_{\rm QCD}$ and a well-defined spectrum of color singlet states.
Its fundamental string dual is unknown. The duality should be extended
to include different boundary
conditions and non conformal quantum field theories.
We may expect that a dual gravitational description would emerge in the strong
coupling regime of QCD. Indeed, the string dual should remain well defined also 
in a highly curved space where the AdS radius become small compared to the
string size\cite{Horowitz:2006ct}. 

In practice, we can deduce salient properties of the QCD dual theory
by studying its general behavior, such as its ultraviolet limit
at the conformal AdS boundary $z \to 0$, as well as 
the large-$z$ infrared region, characteristic of strings 
dual to confining gauge theories.  The fifth
dimension in the anti-de Sitter metric corresponds to the scale transformations of the quantum
field theory, thus incorporating the renormalization group flow of
the boundary theory.  This approach,
which can be described as a bottom-up approach, has been successful
in obtaining general properties of scattering processes of QCD
bound states\cite{Polchinski:2001tt,Brodsky:2003px}, the
low-lying hadron spectra\cite{deTeramond:2005su,Erlich:2005qh},
hadron couplings and chiral symmetry
breaking\cite{Erlich:2005qh,Hong:2005np}, quark potentials in confining
backgrounds\cite{Boschi-Filho:2005mw} and pomeron
physics\cite{Boschi-Filho:2005yh}.

In contrast to the simple bottom-up approach described above, 
a top-bottom approach consists in studying the full supergravity 
equations to compute the glueball
spectrum\cite{Witten:1998zw} or the introduction of additional higher
dimensional branes to the ${\rm AdS}_5 \times {\rm S}^5$
background\cite{Karch:2002sh}, as a prescription for the introduction
of flavor with quarks in the fundamental representation and the
computation of the meson spectrum.

It has been shown recently that the string amplitude $\Phi(z)$
describing  hadronic modes in $\rm{AdS}_5$ can be precisely mapped to
the light-front wavefunctions $\psi_{n/h}$ of hadrons in physical
space-time\cite{Brodsky:2006uq}. Indeed, there is 
an exact correspondence between the
holographic variable $z$ and an impact variable $\zeta$ which
represents the measure of the transverse separation of the constituents 
within the hadrons. This remarkable holographic
feature follows from the fact that current matrix element in AdS
space can be mapped to the exact Drell-Yan-West formula at the asymptotic
AdS boundary\cite{Brodsky:2006uq}. It was also found that effective Schr\"odinger equations
describing hadronic bound states can be
expressed as $3+1$ QCD light-front wave equations\cite{Brodsky:2006uq}.

The boost invariant light-front wavefunctions (LFWFs) in the Fock
expansion at fixed
light-cone time $x^+ = x^0 + x^3$ of any hadronic 
system $\widetilde\psi_{n/h}(x_i,\mathbf{b}_{\perp i}, \lambda_i)$, encode all its
bound-state quark and gluon properties and their behavior in
high-momentum transfer reactions\cite{Brodsky:2004tq}. 
The light-cone momentum fractions $x_i = k^+_i/P^+$ and 
the impact position variables ${\mathbf{b}_{\perp i}}$ represent the relative coordinates of 
constituent $i$ in Fock state $n$, and $\lambda_i$ the helicity along
the $\mathbf{z}$ axis.

In the case of a two-parton constituent bound state the correspondence 
between the string amplitude $\Phi(z)$ and the light-front wave
function $\widetilde\psi(x,\mathbf{b})$ is expressed in closed form\cite{Brodsky:2006uq}
\begin{equation}  \label{eq:Phipsi}
\left\vert\widetilde\psi(x,\zeta)\right\vert^2 =
\frac{R^3}{2 \pi} ~x(1-x)~ e^{3 A(\zeta)}~
\frac{\left\vert \Phi(\zeta)\right\vert^2}{\zeta^4},
\end{equation}
where $\zeta^2 = x(1-x) \mathbf{b}_\perp^2$. 
The variable $\zeta$, $0 \le \zeta \le \Lambda^{-1}_{\rm QCD}$, represents the invariant
separation between point-like constituents, and it is also the
holographic variable $z$ in AdS; {\it i.e.}, we can identify $\zeta = z$.
In the ``hard wall'' approximation\cite{Polchinski:2001tt} the nonconformal metric factor
$e^{3A(z)}$ is a step function. 

\begin{figure}[ht]
\centering
\includegraphics[angle=0,width=8.2cm]{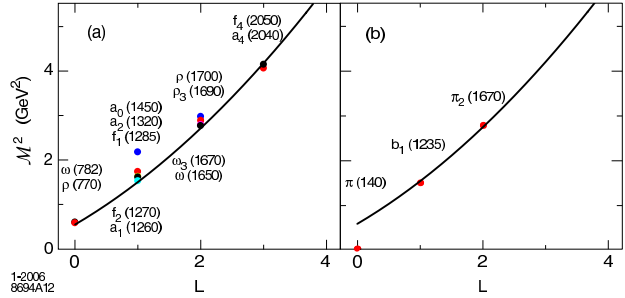}
\caption{Light meson orbital states for $\Lambda_{\rm QCD}$ = 0.32 GeV:
(a) vector mesons and (b) pseudoscalar mesons.}
\label{fig:MesonSpec}
\end{figure}

The short-distance behavior of a hadronic state is
characterized by its twist  (dimension minus spin) 
$\tau = \Delta - \sigma$, where $\sigma$ is the sum over the constituent's spin
$\sigma = \sum_{i = 1}^n \sigma_i$. Twist is also equal to the number of partons $\tau = n$.
Matching the boundary behavior of string modes $\phi(z)$ with the twist of the
boundary interpolating operators we find, upon the substitution
$\phi(z) = z^{-3/2} \Phi(z)$ 
in the  
wave equations in AdS space, an effective Schr\"odinger equation as a function of the
weighted impact variable $\zeta$
\begin{equation} \label{eq:Scheq} 
\left[-\frac{d^2}{d \zeta^2} + V(\zeta) \right] \phi(\zeta) = \mathcal{M}^2 \phi(\zeta),
\end{equation}
with the effective potential
$V(\zeta) \to - (1-4 L^2)/4\zeta^2$ in the conformal limit\cite{Brodsky:2006uq}. The solution
to (\ref{eq:Scheq}) is 
$\phi(z) = z^{-\frac{3}{2}} \Phi(z) = C z^\frac{1}{2} J_L(z\mathcal{M})$. 
Its lowest stable state is determined by the Breitenlohner-Freedman 
bound\cite{Breitenlohner:1982jf}. Its
eigenvalues are obtained from the boundary conditions at $\phi(z =
1/\Lambda_{\rm QCD}) = 0$,
and are given in terms of the roots of the Bessel functions: 
$\mathcal{M}_{L,k} = \beta_{L,k} \Lambda_{\rm QCD}$.
Normalized LFWFs $\widetilde\psi_{L,k}$ follow from
(\ref{eq:Phipsi})
\begin{equation} 
\widetilde \psi_{L,k}(x, \zeta) 
=  B_{L,k} \sqrt{x(1-x)} 
J_L \left(\zeta \beta_{L,k} \Lambda_{\rm QCD}\right) 
\theta\big(z \le \Lambda^{-1}_{\rm QCD}\big),
\end{equation}
where $B_{L,k} = \pi^{-\frac{1}{2}} {\Lambda_{\rm QCD}}/ J_{1+L}(\beta_{L,k})$.
The spectrum of the light mesons is compared in Figure
\ref{fig:MesonSpec} with the data listed by the PDG\cite{Eidelman:2004wy}.

A different approach consists on matching AdS results following Migdal procedure
for the regularization of UV conformal correlators, using 
Pad\'e approximants to build the spectrum with poles of zeros of
Bessel functions\cite{Migdal:1977nu}. This has been discussed recently for two-
and three-point functions\cite{Erlich:2006hq}.

\begin{figure}[ht]
\centering
\includegraphics[angle=0,width=10.6cm]{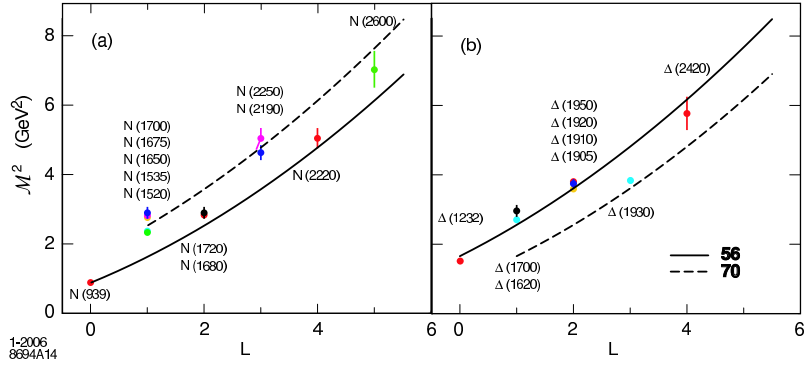}
\caption{Predictions for the light baryon orbital spectrum for
$\Lambda_{QCD}$ = 0.25 GeV. The  $\bf 56$ trajectory corresponds to 
$L$ even  $P=+$ states, and the $\bf 70$ to $L$ odd  $P=-$ states.
The only parameter is the value of $\Lambda_{\rm QCD}$ which is fixed 
by the proton mass. }
\label{fig:BaryonSpec}
\end{figure}

Consider the twist-three, dimension $\frac{9}{2} + L$,
baryon operators $\mathcal{O}_{(9/2) + L} =  \psi D_{\{\ell_1} \dots
D_{\ell_q } \psi D_{\ell_{q+1}} \dots D_{\ell_m\}} \psi$. Since we are taking a
product of operators at the same point, we match the  
dependence of the corresponding AdS  
spin-$\frac{1}{2}$ or $\frac{3}{2}$ modes to the boundary
operators at the ultraviolet $Q \to
\infty$ or $z \to 0$ limit.  
A three-quark baryon is described by wave equation\cite{Brodsky:2006ip}
\begin{equation} 
\left[z^2~ \partial_z^2 - 3 z~ \partial_z
+ z^2 \mathcal{M}^2 - \mathcal{L}_\pm^2 + 4\right] \psi_\pm(z) = 0 
\end{equation}
with $\mathcal{L}_+  = L + 1$, $\mathcal{L}_- = L + 2$, and solution 
\begin{equation} \label{eq:DiracAdS}
\Psi(x,z) = C e^{-i P \cdot x}  
\left[\psi(z)_+~ u_+(P)
+ \psi(z)_-~ u_-(P) \right],
\end{equation}
with $\psi_+(z) = z^2 J_{1+L}(z \mathcal{M})$ and $\psi_-(z) = z^2
J_{2+L}(z \mathcal{M})$.
The constant $C$ in (\ref{eq:DiracAdS}) is determined by the
normalization $R^3 \int  \frac{dz}{z^3}  
\frac{1}{2}\left[\vert \psi_+(z)\vert^2 +
  \vert\psi_-(z)\vert^2\right]=1$ and is given by
$C = \sqrt{2} R^{-\frac{3}{2}} \Lambda_{\rm QCD} / J_0(\beta_{1,1})$.
The physical string solutions have plane waves and chiral spinors
$u_\pm(P)$ along the Poincar\'e coordinates and hadronic invariant
mass states $P_\mu P^\mu = \mathcal{M}^2$. Similar solutions
follow from the Rarita-Schwinger AdS modes $\Psi^\mu$ in the $\Psi_z =
0$ gauge. In the large
$P^+$ limit $\psi_\pm$ are the light-cone $\pm$ components along the 
$\mathbf{z}$ axis: $\psi_+ = \psi^\uparrow$, $\psi_- =
\psi^\downarrow$.  The four-dimensional spectrum follows from
$\psi_\pm(z=1/\Lambda_{\rm QCD}) = 0$:
$\mathcal{M}_{\alpha, k}^+ = \beta_{\alpha,k} \Lambda_{\rm QCD}, ~~
\mathcal{M}_{\alpha, k}^- = \beta_{\alpha + 1,k} \Lambda_{\rm QCD}$,
with a scale independent mass ratio\cite{deTeramond:2005su}.
Figure \ref{fig:BaryonSpec}(a) shows the predicted orbital spectrum of the
nucleon states and Fig. \ref{fig:BaryonSpec}(b) the $\Delta$ orbital
resonances. The data is from [\refcite{Eidelman:2004wy}]. 
The internal parity of states is determined from the SU(6)
spin-flavor symmetry.

The predictions for the lightest hadrons are improved 
relative to the results of [\refcite{deTeramond:2005su}] using the
boundary conditions determined by twist instead of conformal dimensions.
The model is remarkably successful in organizing the hadron spectrum, although
it underestimates the spin-orbit splittings of the $L=1$ states. A better understanding
of the relation between chiral symmetry breaking and confinement is required to
describe successfully the pion. This would probably need a description
of quark spin-flip mechanisms at the wall.

\begin{figure}[ht]
\centering
\includegraphics[angle=0,width=5.3cm]{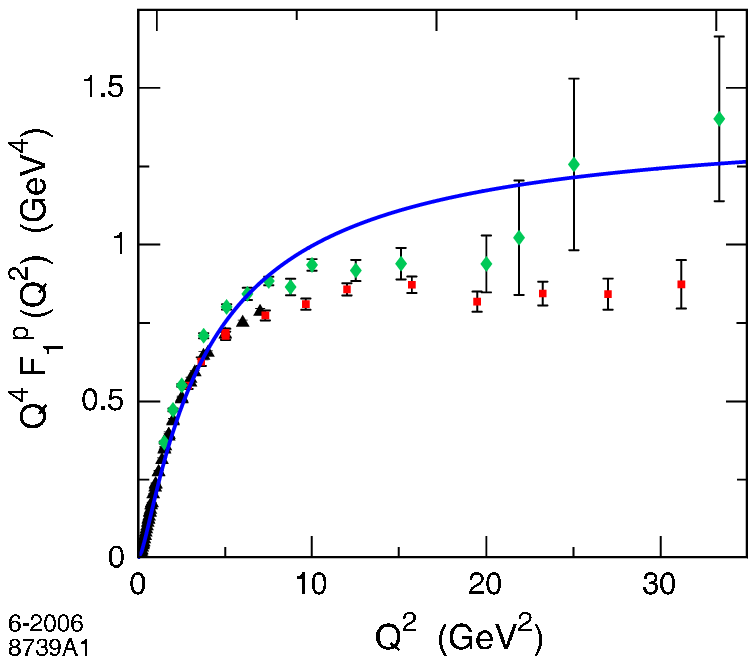}
\includegraphics[angle=0,width=5.5cm]{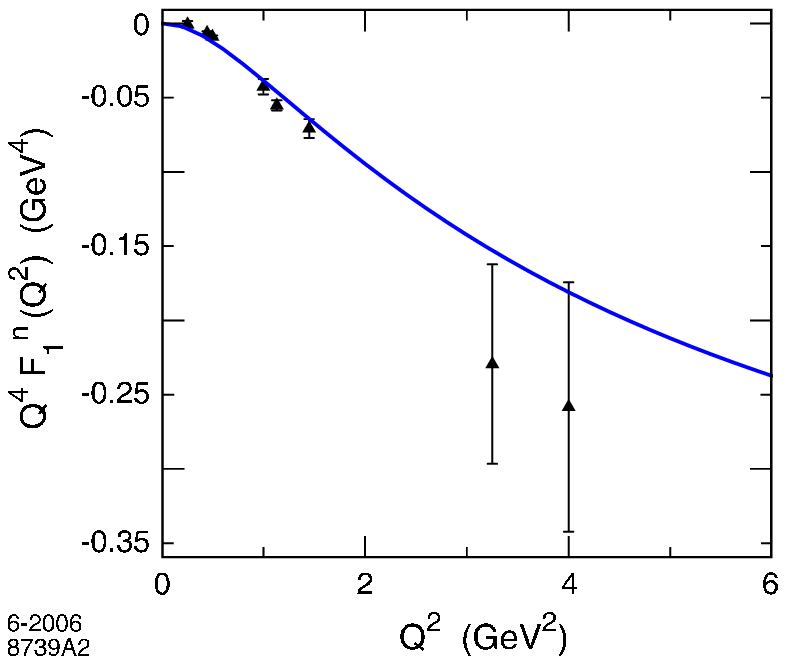}
\caption{Prediction for $Q^4 F_1^p(Q^2)$ and $Q^4 F_1^n(Q^2)$ in the valence
approximation  for 
$\Lambda_{\rm QCD} =  0.21$ GeV. Analysis of the data is from 
Diehl. Data from Sill (solid boxes in red) and superimposed data from
Kirk (solid diamonds in green).}
\label{fig:nucleonFF}
\end{figure}

We now consider the spin non-flip nucleon form factors in the hard wall
model. The effective charges are
determined from the spin-flavor structure of the theory.
We choose the struck quark to have $s^z=+1/2$. The two AdS solutions $\psi_+$ and $\psi_-$
correspond to nucleons with $J^z= +1/2$ and $-1/2$.
For $SU(6)$ spin-flavor symmetry\cite{Brodsky:2006ip}  
\begin{eqnarray}
F_1^p(Q^2) &=&  R^3 \int  \frac{dz}{z^3} \, J(Q,z) 
\vert \psi_+(z)\vert^2 ,\\
F_1^n(Q^2) &=& - \frac{1}{3}  R^3 \int  \frac{dz}{z^3} \, J(Q,z)
\left[\vert \psi_+(z)\vert^2 - \vert\psi_-(z)\vert^2\right], 
\end{eqnarray}
where $J(Q,z)$ is a solution to the AdS wave equation for the external
electromagnetic current polarized along the Minkowski coordinates,
$A_\mu = \epsilon_\mu e^{-i Q \cdot x} J(Q,z)$, $A_z = 0$,
subject to the boundary conditions  $J(Q = 0,z) = J(Q,z=0)=1$ and is 
given by $J(Q,z) = z Q K_1(z Q)$\cite{Polchinski:2001tt}.
The conditions  $F_1^p(0) = 1$ and $F_1^n(0) = 0$ follow
from the identity
$\int_0^1 x dx \left[J_\alpha^2(x \beta) - J_{\alpha+1}^2(x \beta)\right] =
J_\alpha(\beta) J_{\alpha+1}(\beta)/\beta$. 
Figure \ref{fig:nucleonFF} compares the predictions for the
Dirac nucleon  form factors with the experimental data\cite{Diehl:2005wq}.

\vspace{-5pt}
 
We have shown how the string amplitude $\Phi(z)$ defined on the fifth dimension in $AdS_5$ 
space can be precisely mapped to the light-front wavefunctions of hadrons 
in physical spacetime\cite{Brodsky:2006uq}. This specific
correspondence provides an exact holographic mapping
in the conformal limit at all energy scales between string modes in
AdS and boundary states with well-defined number 
of partons. Consequently, the AdS string mode $\Phi(z)$ can be regarded as the probability
amplitude to find $n$-partons at transverse impact separation $\zeta =
z$. Its eigenmodes determine the hadronic mass spectrum. 
Although major dynamical questions remain to be solved for extending
the duality from large to small 't Hooft coupling,  the
string-parton correspondence described in [\refcite{Brodsky:2006uq}] 
suggests that basic features of QCD can be understood in terms
of a higher dimensional dual gravity theory 
which holographically encodes multi-parton boundary states into string modes and 
allows the computation of physical observables at strong coupling.

\vspace{5pt}

\noindent{\bf Acknowledgements}

This work was done in collaboration with Stan Brodsky. We thank
Misha Shifman for his invitation to CAQCD 2006 and Joe
Polchinski for encouraging comments.

\end{document}